# Pseudoscopic imaging in a double diffraction process with a slit: critical point properties

**José J. Lunazzi and Noemí I. Rivera**

*Universidade Estadual de Campinas, Instituto de Física Gleb Wataghin, Caixa Postal 6165, 13083-852 Campinas, SP, Brazil*

Pseudoscopic (inverted depth) images that keep a continuous parallax were shown to be possible by use of a double diffraction process intermediated by a slit. One diffraction grating directing light to the slit acts as a wavelength encoder of views, while a second diffraction grating decodes the projected image. The process results in the enlargement of the image under common white light illumination up to infinite magnification at a critical point. We show that this point corresponds to another simple-symmetry object-observer system. Our treatment allows us to explain the experience by just dealing with main ray directions.

# 1. INTRODUCTION

Direct pseudoscopic images are uncommon in optics. Recently, a pseudoscopic image was demonstrated in refractive optics,[1] but more widely known are registered images, such as inverted stereo pairs or holographic images. Pseudoscopic holographic imaging may render inverted depth images in continuous parallax exclusively under monochromatic light or through some process that renders the final image monochromatic at least over the horizontal field of view. We demonstrated previously,[2] that diffraction could be combined with a simple process to obtain reversed depth images whose continuous horizontal parallax is due exclusively to diffractive elements. This new kind of image is of interest because it opens up the possibility of comparing the its parallax field and image quality with those produced by conventional imaging systems. New optical systems based on double diffraction intermediated by a refractive element[3] could also benefit from knowledge of the properties of a simpler system such as ours. In this paper we show additional properties of this kind of image: image resolution, presence of astigmatism, viewing angular field, field of view, and magnification. We analyze the case of propagation in a plane that is not perpendicular to the gratings, giving experimental evidence of the pseudoscopic image, its spectral composition, its magnification and the existence of a critical point.

The system is depicted in the simple symmetry of Fig. 1. Our system consists of two identical diffraction gratings, *DG1* and *DG2*, symmetrically located at either side of an aperture *a*. The plane of the figure corresponds to the horizontal plane $P$ of the aperture, while the lines of the grating are in a vertical direction. $\lambda_M$, $\lambda_m$ are the extreme wavelengths of the visible spectrum. An object of white or gray tonality is illuminated by common white light diffusing at a very wide angle, for example, from point *A*, whose coordinates are *X, Y, Z*. $X_i$ is the generic coordinate corresponding to the point of incidence of the light rays that impinge on the first grating and may reach the slit after diffraction.

We consider the part of the beam reaching the grating that, after diffraction, travels toward the aperture. $Z_R$ is the distance from the slit to the first grating, made equal to the distance from the slit to the second grating, the symmetry condition. Deviation of rays by diffraction is given by the basic grating equation,

$$\sin\theta_i - \sin\theta_d = \lambda v, \qquad (1)$$

where $\theta_i$ represents the angle of incidence of light traveling from point $A$ to points on the grating, $\theta_d$ represents the angle of diffraction of light that travels from points on the grating to point $P$, $v$ is the inverse of the grating period, and $\lambda$ represents the wavelength value corresponding to each ray.

We see from Fig. 1 and Eq. (1) that the angle $\theta_i$ of incidence into the first grating through which the wavelength $\lambda$ will be allowed to enter the slit can be obtained from the equation

$$\theta_i = \arcsin\left[\frac{(X - X_i)}{\sqrt{(X - X_i)^2 + Z^2}}\right], \qquad (2)$$

where the value of $X_i$ is known from the relationship

$$\frac{(X - X_i)}{\sqrt{(X - X_i)^2 + Z^2}} - \frac{X_i}{\sqrt{X_i^2 + Z_R^2}} = \lambda v. \qquad (3)$$

This is the way in which we demonstrated the formation of the pseudoscopic image. In Fig. 1, we are dealing with an image that is convergent in the horizontal plane but that does not converge vertically due to beam divergence in the vertical plane. This situation changes when we consider an observer, because the reduced diameter of the pupils of the observer's eyes causes the image to be convergent on the retinas.

## 2. IMAGE RESOLUTION

The perfect symmetry of the system precludes the existence of aberrations in a horizontal ray tracing analysis so that is not appropiated to make the classical first order and third order aberration analysis. The image quality can only be degraded horizontally due to the width of the slit or the defects of the diffracting elements. The slit does not interfere in the direct visual observation process vertically, but it creates a lateral wavelength blurring of the same angular extension as its aperture similar to that of a pinhole camera. We see the situation in Fig. 2, where the width of the slit was effectively considered by including the two rays that pass through the borders of the slit instead of a single ray going through its center. Two points $A'_1$ and $A'_2$ are the symmetricals of the object point A corresponding to the border points $P_1$ and $P_2$, defining an extension $2a$. The dashed lines were added to the scheme to help understanding this. It appears to the observer that the image point has an extension $c = 2a\cos\theta_i$.

## 3. PRESENCE OF ASTIGMATISM AND CONSIDERATIONS ON IMAGE QUALITY

It happens that the eye of the observer will always be focusing at points whose distance corresponds to the distance traveled by the light in the vertical plane, and at the same time, at the points where the slit and the second diffraction grating perform the image, in the horizontal plane. This combination of two focal properties creates astigmatism (see Fig. 3) which is proportional to the diameter of the eye's pupil. Calculating with the Emsley[4] model of the eye and considering a pupil diameter of 3 mm and the data of our experiment, the astigmatic image of a point results 0.07 mm high at the retina.

The perception of depth by binocular vision is a triangulation operation acting on the horizontal plane, so that depth is determined by the horizontal rays. We can highlight the particular properties of our system having reduced chromatic aberrations because of the wavelength spreading distribution. Surface flaw of the diffractive elements are also less important due to the tolerance of transmission gratings, which makes deflection be dependent mainly on the period of the grating and very little on the inclination of the grating. Geometric distortion exists only as the result of a different vertical and horizontal scale, linearity being preserved on each of both scales. They can eventually be matched by properly choosing the value of the period for both gratings.

## 4. VIEWING ANGULAR FIELD

The horizontal angular parallax can be calculated by considering two points of the object and the path that is allowed to light whose wavelength values are on extremes of the spectrum. The angular region limited by this rays is the common region for simultaneous observation of the two points and includes all intermediary points (see Fig. 4). $X_{Ai}$, $X_{Bi}$ are coordinates corresponding to the points of incidence of the light rays from points *A* and *B* respectively in the first grating. Due to the symmetry of the system all rays reach the second grating at points such as $X_{Ai}'$, $X_{Bi}'$. Employing Eq. (3) to get the value of Xi and introducing it into Eq. (2) the result is:

$$\theta_i = \theta_i(X, Z, Z_R, v, \lambda) \tag{4}$$

An equation which we do not calculated but solved numerically. We can then obtain the angular field $\Delta\theta$ by calculating two $\lambda$ values at the extreme wavelengths at the corresponding extreme lateral points of the object and subtracting them. The analysis must then be made for points at extreme lateral and depth points of the object. The resulting expression is:

$$\Delta\theta = \theta_i(X, Z, X_B, Z_B, Z_R, v, \lambda_M) - \theta_i(X, Z, Z_A, Z_A, Z_R, v, \lambda_m) \tag{5}$$

We solve it numerically by means of the software Mathematica 4.1. The calculated value for the data of our experimental case is in Table 1 altogether with the experimental result. It is possible to obtain a case where the expression for $\Delta\theta$ is simple by choosing the case which considers point A in a position where $\lambda_M$ is diffracted perpendicullarly through DG1 so impinging into DG2 and point B in a position where $\lambda_m$ does the same. Considering that Eq. (2) can then be substituted for:

$$\theta_i = arcsin\lambda\nu \qquad (6)$$

The symmetry of the problem allows using the same values of the field $\Delta\theta$ at the second stage for analyzing the viewing situation. So that we have:

$$\Delta\theta = arcsin\lambda_M\nu - arcsin\lambda_m\nu \qquad (7)$$

as the value of the horizontal angular parallax. For $\lambda_M$ = 650 nm and $\lambda_m$ = 450 nm we obtain $\Delta\theta = 6.5°$. Doubling the number of lines per millimeter this field is also doubled, indicating that it is possible to make a system for binocular observation of the image.

## 5. FIELD OF VIEW AND CRITICAL POINT

Figure 5 shows as a shadowed angle the object field of view for a fixed position $(x_c, z_c)$ of the observer. CP is the point symmetrical to the observer's position. Assuming two light rays which may go through this point and reach the observer's position, $X_{CPi1}$ and $X_{Cpi2}$ are their abscisas. The region seen of an object depends on the object distance in a manner which resembles that of an object under a converging lens: it can change dramatically when close to a certain critical distance which renders infinite magnification. This happens at the critical point CP as follows: by including in the system an object in three different depth positions $O_1$, $O_2$, $O_3$, we can notice that a very small area is imaged as the object approaches the critical point. The situation is explained in Fig. 5(b) in the case when the viewpoint is symmetric to the object. It is clear from the figures how magnification may vary with longitudinal object position. The whole field of view between $X_2'$ and $X_1'$ can be fulfilled either by some extension of the object or by a single object point in the critical case when the object is at CP. Comparing Fig. 5(a) with Fig. 5(c), the crossing of rays which corresponded to laterally extreme object points explains the lateral inversion on the image.

## 6. CASE OF PROPAGATION IN A PLANE THAT IS NOT PERPENDICULAR TO THE GRATINGS

Our first analysis was based on Fig. 1 and Fig. 2 assuming that the observer is located on the same horizontal plane than the object and it includes a perpendicular to the gratings as shown for observer $O_1$ in Fig. 6. That horizontal plane is then the plane **H** of wavelength dispersion . If we consider now the observer being located over or under that plane the difference would be in that case that dispersion does not happens perpendicularly to the gratings. Let us analize now the propagation in another plane at an angle **α** to plane **H** for the image to be seen by the vertically displaced observer $O_2$. All rays that diverge in that upward direction do not properly satisfy Eq. (1), but another one[5] that includes the value of its angle of incidence as determined by polar coordinates ($\theta, \varphi$).

$$sin\theta_i - sin\theta_d = \frac{\lambda \nu}{\sqrt{1-\varphi^2}} \qquad (8)$$

This diffraction case is denominated as "conical diffraction" by some authors[6]. Although a ray may rise over the oblique plane it goes under it after the slit and finally the point-symmetric situation remains the same. A new plane of symmetry could be drawn for each ray which enters at any azimuthal angle but the image point **i** can not be precisely determined anymore, and some aberration may be present. The vertical parallax and field of view are easy to predict by triangulation because both are directly allowed by the height of the slit.

## 7. INCREASING THE DISTANCE TO THE SECOND GRATING

When the distance of the second grating to the slit is doubled, because the angles of incidence at DG2 are the same, two consequences are clear: the image is magnified because it is extended through the grating and the observer must move to $(x_c', z_c')$ to keep the critical point at the same place than before (see Fig. 7). If the observer moves laterally to follow the view of the object while the grating is being displaced, the result will be that of Fig. 8 and the position of the critical point comes closer to the system.

## 8. EXPERIMENTAL RESULTS

We employed two plastic embossed holographic transmission gratings of the same type, commercially available for architectural or educational purposes. Both had $533 \pm 5$ lines/mm and were fixed between two glass plates 2 mm thick. Their effective area employed was 60 mm x 40 mm. They were located $600 \pm 2$ mm apart in a parallel position and a vertical black paper slit $0.7 \pm 0.15$ mm wide was in between both gratings. Parallelism of the grating planes was verified to better than $\pm 1$ mm by making coincident reflections of a diode laser beam which traversed the slit, impinged on both gratings and returned to the laser exit window. Photographs were made with SONY video 8 Handycam camera connected to a Pentium I computer to get 240 x 320-pixel resolution. In a first experience we employed an 85 mm long almost vertical luminous filament from a 300 W halogenous lamp, inclined with its higher extreme being at a distance of 230 mm to the grating, closer than its other extreme which was at 280 mm from it. The camera was at a 1,015 mm distance. The object was inclined frontally in order to show with a stereo photographic pair that the image is pseudoscopic. We found two viewpoints showing the filament in a frontal view and in a lateral view, demonstrating the depht inversion in Fig. 9 because the right-hand side view shows a vertical line while the left-hand side view shows the upper part inclined to the left, a parallax sequence corresponding to an object which is inclined to the back. A second experience with the object at the same position showed the image seen from the same viewpoint (red wavelength) as shown in Fig. 10 where we show the unfiltered image to be compared with the image as seen alternatively with two interference filters. The wavelength limitation shows the correspondance between wavelength value and the horizontal part of the image. Our third experiment employed as object an halogeneous 50 W lamp with its mirror and compares the original image with the image obtained after doubling the distance of the second grating to the slit. The object was located at X = 221 ± 5 mm , Z = 68 ± 5 mm and observer at $x_c = 202 \pm 5$ mm and $z_c = 928 \pm 5$ mm The observer's viewpoint needed to be changed 80 mm to the right. The image is seen in Fig. 11 and resulted magnified by a factor of 4. In our fourth experiment we employed as object the shadow of a paper clip located against a diffusing background. We needed to do that instead of illuminating the clip itself because we needed a brighter image on the camera to be accepted by the exposure system. The average mean intensity in this case results to be much higher. The clip is made of wire 1 mm thick and we employed its asymmetric shape to identify the inversion properties corresponding to the dispersion direction. The background remained fixed at 1,680 mm distance while the clip was located at three different positions: between the system and the critical point at 960 mm, at the critical point at 1,270 mm, and farther from the critical point at 1,540 mm, respectively. The situation is illustrated in Fig. 12 and the resulting images in Fig. 13. Besides showing clearly the inversion the image appears quite undistorted and infinite magnification is evident through the whole extension of the elongated image resulting at the critical point. Our final experiment consisted in measuring the viewing angular field. For simplicity of comparison the result was included with the theoretical values of Table 1.

## 9. CONCLUSIONS

The symmetry of the developed system shows an easy way to analyze its images, and the quality of the images is close to the resolution of the naked eye. Although not intended for educational purposes but for developing new imaging systems having less weight and volume, the system is so simple that any person with minimal skill can experience the new properties of diffractive white light imaging. We may then conclude that the ray distribution generated after the second diffraction may be always considered as a decoding of the effect of the primary grating and that the original intensity distribution of rays from the object was maintained through this selective wavelength distribution process. Although the diffracting elements acts in one direction only and does not project an image, it is possible to have a two-dimensional equivalent by using circular diffraction gratings, a result to be published elsewhere.

## ACKNOWLEDGEMENTS

The "Pro-Reitoria de Pós Graduação" of Campinas State University-UNICAMP is acknowledged for a fellowship for miss. Noemí Inés Rodríguez Rivera. The "Fundo de Assistência ao Ensino e à Pesquisa – FAEP" of Campinas State University-UNICAMP is acknowledged for financial support. Paul and Silvia Baldi are acknowledged for reviewing the English grammar.

The authors can be reached by e-mail as follows: José J. Lunazzi, lunazzi@ifi.unicamp.br, and Noemí I. Rivera, nrivera@ifi.unicamp.br.

# TABLES

**Table 1. Experimental checking of the angular field. $\Delta\theta_T$ (º) is the theoretically alculated value while $\Delta\theta_E$ (º) is the measured value.**

---

$\lambda_A = 500\,nm \quad \lambda_B = 670\,nm$

---

| $x_A$ (mm) | $x_B$ (mm) | $z_R$ (mm) | $\Delta\theta_T$ (º) | $\Delta\theta_E$ (º) |
|---|---|---|---|---|
| 28 | 17 | 300 | 2.7 | 2.4 |

| $z_A$ (mm) | $z_B$ (mm) | | | |
|---|---|---|---|---|
| 209 | 179 | | | |

---

# FIGURE CAPTIONS

Figure 1. Ray-tracing scheme for the depth inverted image.

Figure 2. Schematic ray tracing showing the limit of sharpness due to the width of the slit.

Figure3. Schematic perspective view of the setup including ray tracing that generates vertical astigmatism. A = point object ; A' = astigmatic image .

Figure 4. Ray-tracing scheme for calculating the angular field of view.

Figure 5. Ray-tracing scheme showing object field of view. Object in three positions around the critical point. ($0_1$) in front ($0_2$) at the critical point. ($0_3$) behind the critical point.

Figure 6. Single ray path for a vertically displaced observer.

Figure 7. Observer's position that keeps the critical point invariant when displacing the second grating.

Figure 8. Ray-tracing scheme showing magnification and observer's position when the second grating's distance is doubled.

Figure 9. Photographic evidence of the pseudoscopy of the image. (a) right-hand side image (blue wavelength) (b) left-hand side image (red wavelength).

Figure 10: (a) above: unfiltered red wavelength view of the filament. (b) same view after filtering to 634-640 nm bandwidth. (c) same view as (a) but filtered to 643-657 nm bandwidth.

Figure 11. Magnification by displacement of the second grating. (a) Undisplaced. (b) Displaced 300 mm.

Figure 12. Schematic view of the experience made to show the properties of the critical point.

Figure 13. Sequence of object positions at increasing distances showing the inversion of the image. (a) between the first grating and the critical point. (b) at the critical position (c) farther from the critical position, showing lateral inversion.

**FIGURES**

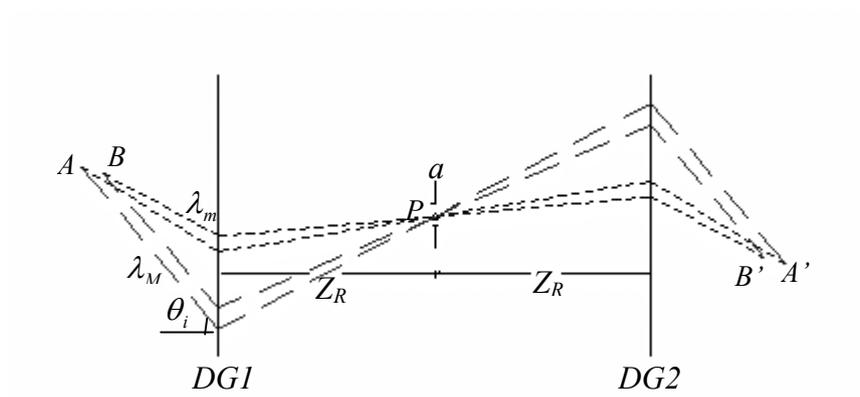

Figure 1.

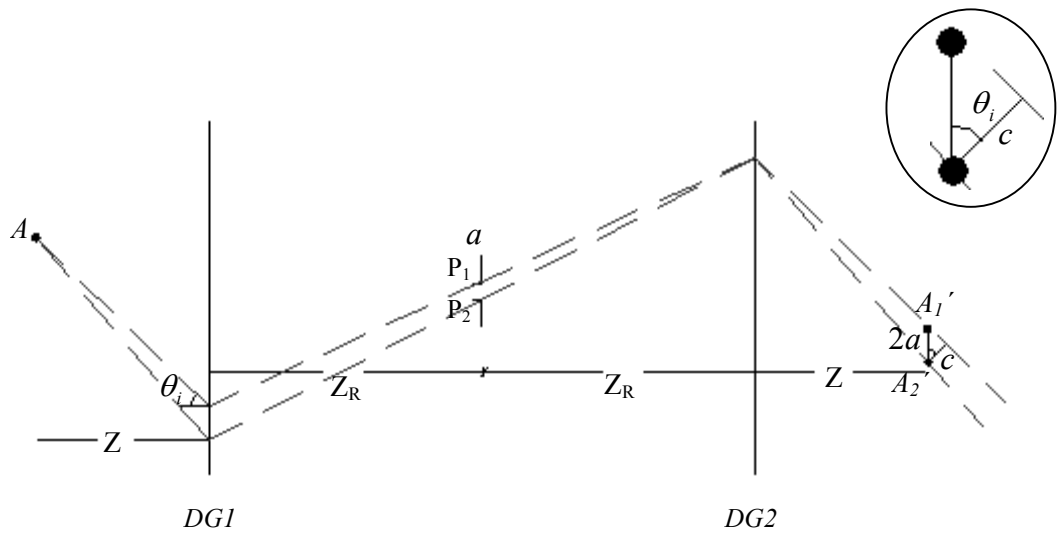

Figure 2.

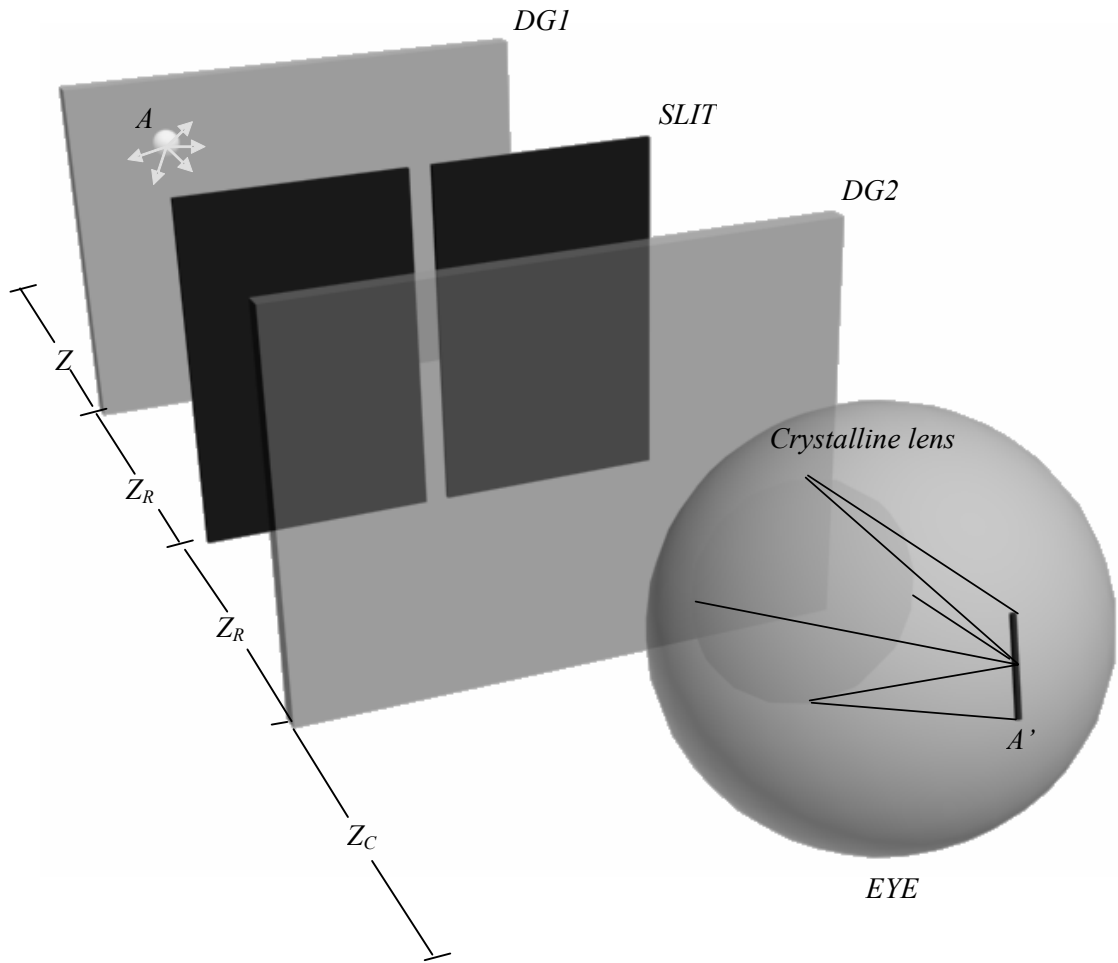

Figure 3.

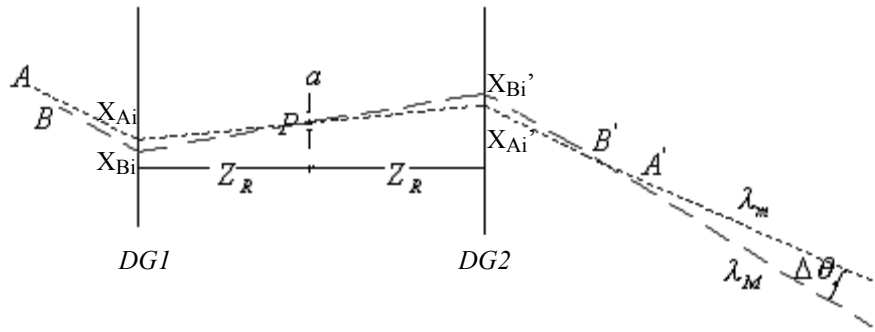

Figure 4.

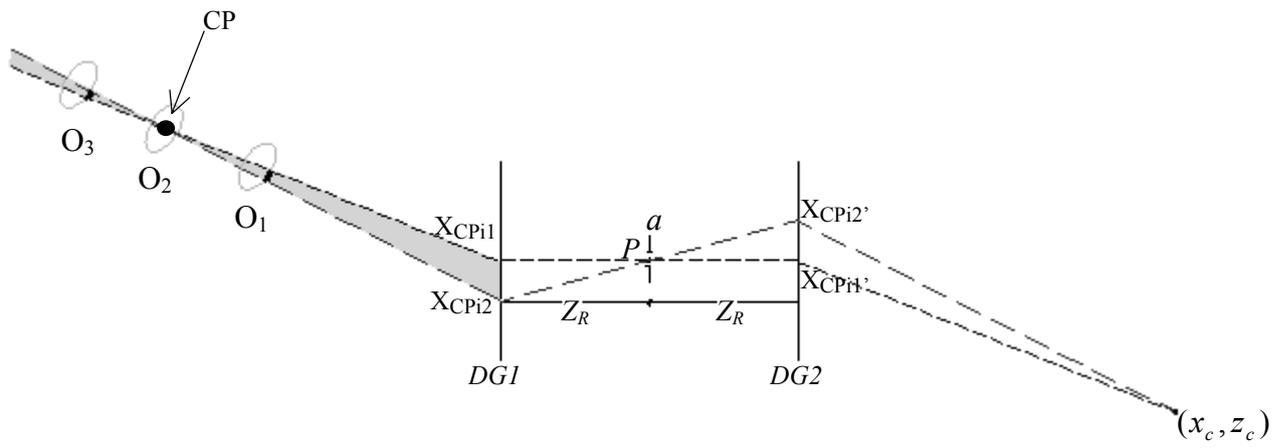

Figure 5.

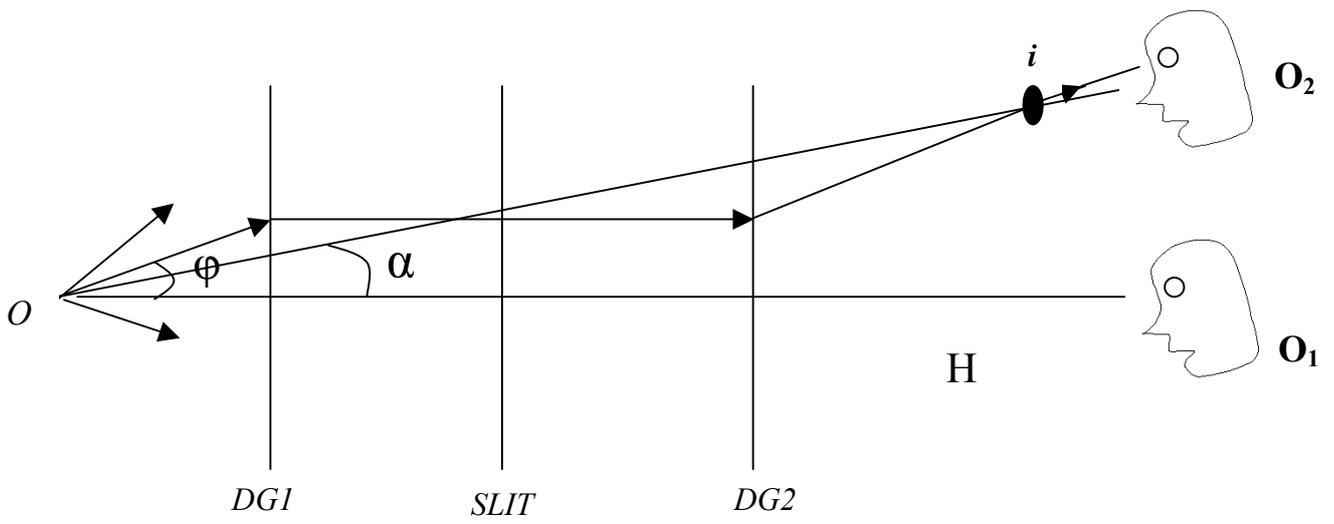

Figure 6.

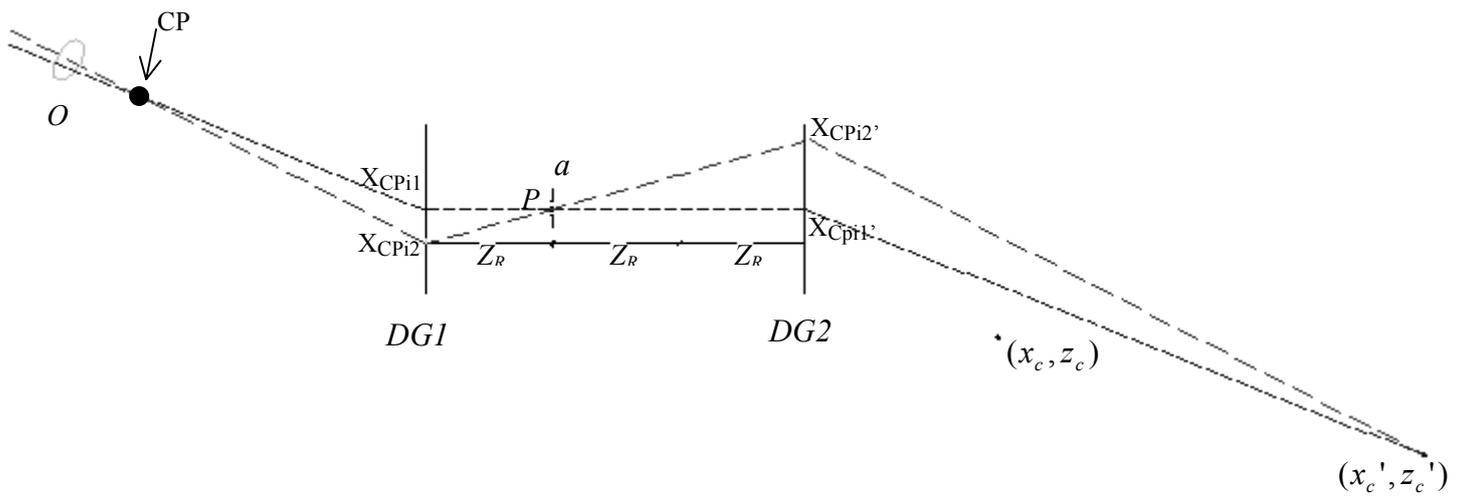

Figure 7.

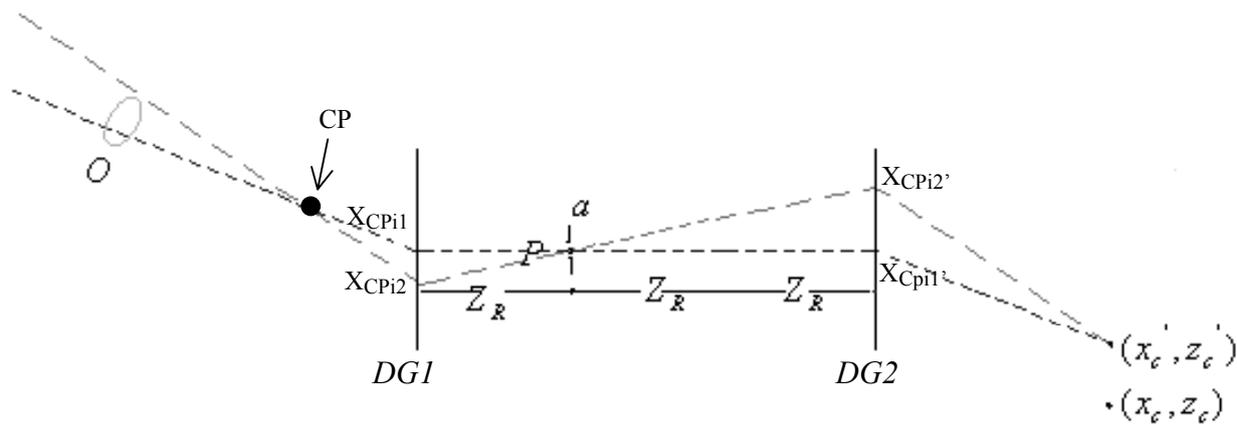

Figure 8.

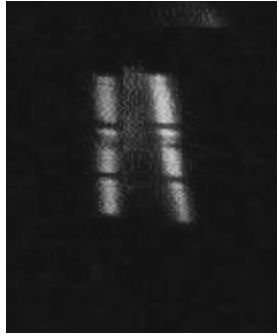

Figure 9.

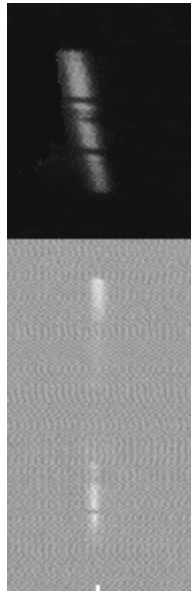

Figure 10.

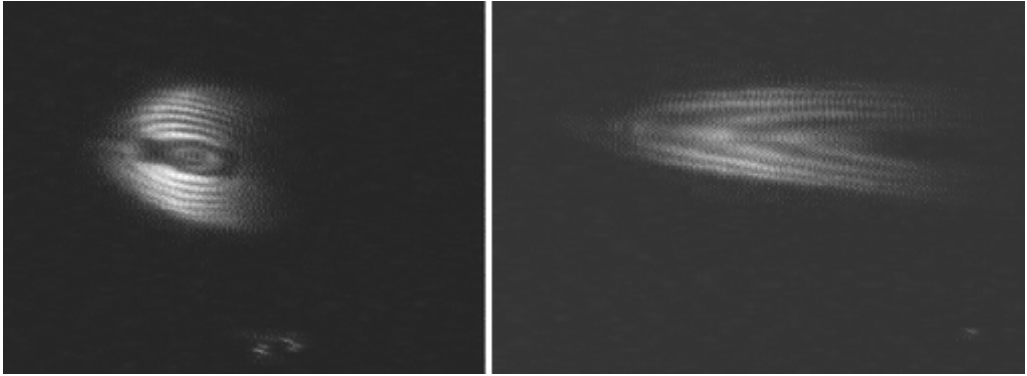

(a)  (b)

Figure 11.

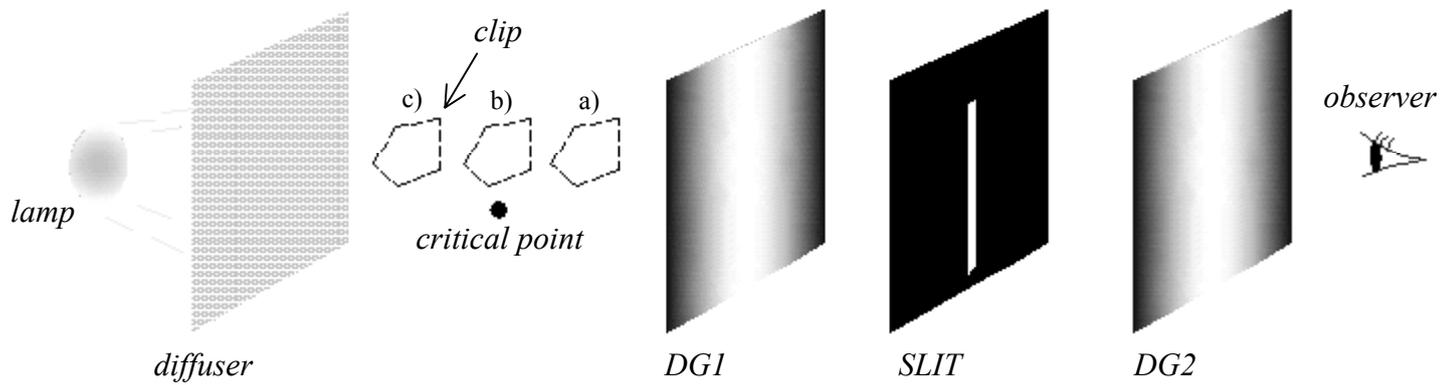

Figure 12.

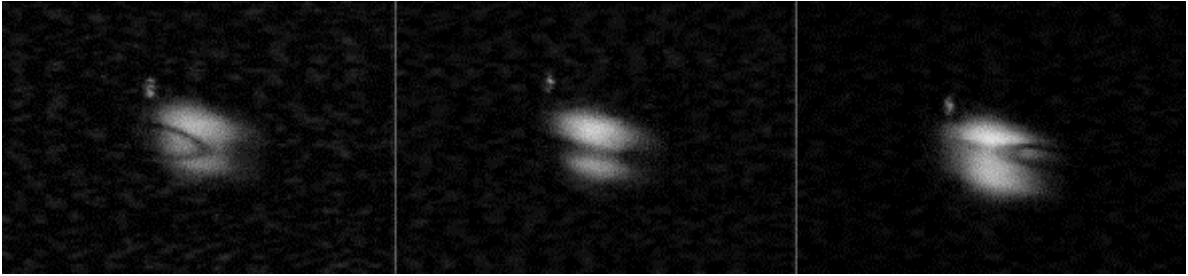

(a) (b) (c)

Figure 13.